\documentclass[11pt]{article}
\include{epsf}

\newcommand{\be}{\begin{equation}}
\newcommand{\ee}{\end{equation}}

\begin{document}
\title{\Large \bf Novel approach for spin-flipping a stored polarized beam
\thanks{Supported by a research grant from the US Department of Energy}} 

\author{Ya.S. Derbenev and V.A. Anferov\\
\it Physics Department, University of Michigan, Ann Arbor, MI48109-1120}    
\date{\today}
\maketitle
\vspace*{-1cm}

\begin{abstract}
The traditional method of spin-flipping a stored polarized beam is based
on slowly crossing an rf induced depolarizing resonance.  This paper
discusses a novel approach where the polarization reversal is achieved by trapping  
the beam polarization into a stable spin-flipping motion on
top of the rf induced resonance at a half-revolution frequency.
\end{abstract}

\section{Introduction}
       
Developing the spin-flipping technique is important for high energy spin experiments
since frequent reversals of the beam polarization can significantly reduce systematic 
errors in an experiment's spin asymmetry measurements. To spin-flip a stored polarized
beam, one can slowly ramp the frequency of an rf magnet (either solenoid or
dipole) through the rf-induced depolarizing resonance. 
This technique was successfully used to spin-flip a polarized proton
beam stored in the IUCF Cooler Ring with and without Siberian Snake~\cite{iucf-flip}.  
While slow resonance crossing rate is
required to achieve good spin-flip efficiency, it also increases the time of 
each spin-flip. Moreover, it makes the spin motion sensitive to weak 
synchrotron sideband or higher-order depolarizing resonances that may occur in the 
vicinity of the rf induced resonance.  

In this paper we discuss another possible way of spin-flipping the beam polarization
by rearranging the stable spin motion in such a way, that the polarization direction 
alternates on every particle turn around the ring without any depolarization.

\section{Stable spin motion at an RF induced resonance}

In a circular accelerator or a storage ring with no Siberian Snakes, the spin vector of each 
particle precesses around the vertical magnetic field of the ring's dipole magnets. 
For a particle moving along the closed orbit, the spin tune $\nu_s$, 
which is the number of spin precessions during one turn around
the ring, is proportional to the beam energy
\be
\nu_s=G\gamma ,
\ee
where $\gamma$ is the Lorentz energy factor and $G=(g-2)/2$  is  
particle's gyromagnetic anomaly. While for protons $G=1.79285$, it is much smaller
for electrons ($G=0.00116$) and deuterons ($G=-0.1426$).  

This vertical spin precession can be perturbed by any horizontal rf magnetic field,
whenever its frequency is in resonance with the spin motion 
\be
f_{RF} = f_{circ} (k\pm\nu_s),
\ee
where $f_{circ}$ is the beam circulation frequency and $k$ is an integer.
Near the rf induced resonance, the spin precession becomes unstable,
which could lead to depolarization of a vertically polarized beam. 
However, the same rf magnetic field establishes a new stable spin direction
in the horizontal plane.  To show this, let us consider the spin motion
in the presence of an rf spin perturbation $\varepsilon\cdot e^{-i\omega\theta}$,
where $\varepsilon$ is resonance strength  and we
assume that the perturbation frequency $\omega=f_{RF}/f_{circ}\pm k$ is close
to the resonance condition of Eq.(2).  The equation of spin motion can be written for
the spinor wave function $\psi$ in the following form~\cite{spinor}, 
\be
\frac{d\psi}{d\theta}=-\frac{i}{2}\left(
\begin{array}{cc} 
G\gamma  &  \varepsilon\cdot e^{-i\omega\theta} \\
\varepsilon^*\cdot e^{i\omega\theta} & -G\gamma \\ \end{array}
\right) \cdot \psi ,
\ee
where $\theta$ is the azimuthal particle coordinate in the ring.
In these notations, the diagonal terms in Eq.(3) represent spin rotation
around the vertical axis, while the off diagonal terms correspond to horizontal
spin perturbation. 
Transforming the above equation into the resonance rotating frame,  
$\psi= e^{-i\omega\theta\sigma_3/2}\cdot\xi$, 
the equation of spin motion becomes
\bigskip
\be
\frac{d\xi}{d\theta} = -\frac{i}{2}\left( 
\begin{array}{cc} 
G\gamma-\omega  &  \varepsilon \\
\varepsilon^* & \omega-G\gamma \\ \end{array}
\right)\cdot \xi
\ee
Note that exactly on top of the rf induced resonance $G\gamma-\omega=0$, 
and the spin precesses around the spin perturbing field $\varepsilon$, 
which rotates with frequency $\omega=f_{RF}/f_{circ}\pm k$ in the laboratory frame.
Thus, the direction along the rotating spin perturbation vector becomes
stable for the spin motion in the presence of an rf perturbation.
Next, we consider how the existence of such a rotating horizontal
stable spin direction can be used for spin-flipping.

\section{Spin-flipping at selected energies in the absence of Siberian snakes}

As we saw in the previous section, an external rf magnetic field, when at
resonance with the spin precession,
creates a horizontal stable spin direction which rotates around the ring
with angular frequency 
\be
w \theta = \left(f_{RF}/f_{circ}\pm k\right) \theta=G\gamma\theta .
\ee
When $G\gamma$ is equal to a half-integer number, the horizontal spin
would rotate by exactly 180 degrees in one turn around the ring. 
Thus, the spin would flip its direction after every turn.
This opens a possibility to organize the spin motion in such a way that
spin would arrive at the experimental section with longitudinal polarization,
whose sign alternates on every turn.  
A practical solution would require installing an rf solenoid at the 
experimental straight section and matching the injected beam polarization with
the longitudinal direction at the experimental section.  One can also 
use radial rf dipole field to create a spin-flipping motion of a horizontal spin.
The stable spin direction would be radial near the rf dipole,
while in a different straight section it will be rotated towards the longitudinal
direction by angle
\be
\phi = G\gamma \theta_{bend}
\ee 
where $\theta_{bend}$ is the orbit bend between the rf dipole and the point of interest.
Note that the spin tune $G\gamma$ should be half-integer for both the rf solenoid and rf 
dipole induced spin-flipping.

\begin{figure}
\centering\mbox{\epsfysize=3.5in\epsffile{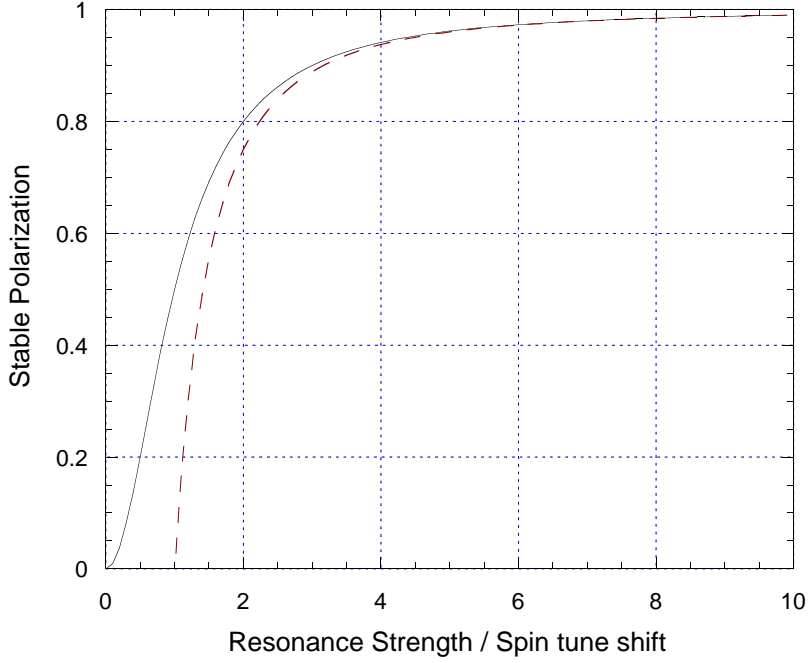}}
\caption{\small The average stable polarization magnitude during the spin-flipping process is plotted
against the $\varepsilon/\Delta\nu_s$ ratio. The solid curve is derived analytically from the spin
transformation over two turns around the ring. The dashed curve is stable polarization predicted by Eq.(8).}
\end{figure}

There are several effects that could potentially perturb the spin
motion driven by the rf magnet.  With an energy offset or spread present in the beam, 
particles will have their spin precession frequency slightly shifted from the
rf resonance.  An offset from the rf resonance would tilt the stable spin 
direction out of the horizontal plane by an angle $\beta$,
\be
\beta = \tan^{-1}\left(\frac{\Delta\nu_s}{\varepsilon}\right) ,
\ee
where $\Delta\nu_s=G\Delta\gamma$ is the spin tune shift due to the energy offset.  
The tilt in the stable
spin direction would in turn reduce the equilibrium horizontal beam polarization $P$
by a factor of
\be
\frac{\Delta P}{P}=1-\cos^2 \beta
\simeq \left(\frac{\Delta\nu_s}{\varepsilon}\right)^2
\ee
Thus, in order to maintain control over the spin motion ($\Delta P/P<0.1$), the rf induced resonance
should dominate over the energy spread,
\be
\varepsilon \ge 3\cdot\Delta\nu_s\ .
\ee
For example, to overcome a spin tune shift of 0.001 the strength of the rf induced resonance
should be about 0.003.  
The average stable polarization magnitude during the spin-flipping process is plotted
against the $\varepsilon/\Delta\nu_s$ ratio in Fig.~1.

Similarly, the rf induced resonance should dominate over higher-order 
horizontal spin perturbations, which would also tend to  perturb the stable
spin direction.  It is also important to note that, in the spin-flipping
method described here, the rf solenoid would operate at exactly one half of the
beam circulation frequency. Therefore, one particular part of the beam 
would always pass the rf magnet when its field is very close to zero,
and thus, would not have stability of the horizontal spin. 
To avoid this problem, one could create a gap in the stored beam while 
synchronizing the rf field with the remaining beam bunches. Another solution 
could be to use a high frequency rf-magnet (or a special rf-cavity) synchronized
with every bunch in the beam. This could be done when there is an odd number of bunches
in the ring; the rf magnet would then be operating at the frequency $f_{RF}=\frac{h}{2}f_{circ}$,
where $h$ is the harmonic number (odd integer) of the main rf cavities.

\section{Spin-flipping in the presence of Siberian snakes}

In the presence of a full Siberian snake~\cite{snk-idea} in the ring, the spin tune becomes half-integer
and energy independent. With a proper choice of the betatron tunes, a set of 
Siberian snakes can overcome all dangerous depolarizing resonances in the ring~\cite{courant-lee}.
Nevertheless, an rf magnetic field in resonance with the half-integer spin tune
can depolarize the beam even in the presence of a Siberian snake.
These resonances are often called "snake" depolarizing resonances~\cite{snk-res}.
Similarly to the case with no snake in the ring, such an rf induced "snake" resonance
can create a rotating stable spin direction.  Since the spin tune is half-integer
in the presence of a Siberian snake, this rotating stable spin direction 
would be flipped after every turn around the ring.

To create such a stable spin-flipping mode of the spin motion, 
the rf magnetic field has to be orthogonal to the unperturbed spin.
With a single snake in the ring and no spin perturbation present, the stable spin
direction is horizontal and coincides with the snake axis
in the straight section opposite to the snake location. 
Therefore, an rf dipole with vertical field operating at $0.5f_{circ}$ would
make vertical direction stable for the spin motion,  
while the polarization direction would flip after every turn around the ring.  

With an even number of snakes in the ring, the unperturbed stable spin direction is 
vertical.  Ideally, one could use longitudinal rf field near the interaction region
to create the longitudinal spin stable and flipping every turn around the ring.
However, solenoids become impractical at high energies since their spin rotation angle
linearly decreases with the beam momentum.
An energy independent spin rotation in a dipole makes dipole magnets more attractive 
for spin manipulation at high energies.
An rf dipole with radial field could create stable spin-flipping of the
beam that has radial polarization near the rf dipole location, while 
at the experimental straight section polarization would be longitudinal. 

The stability of the continuous spin-flipping motion 
could be lost when the spin-tune moves out of the induced rf resonance.
Such a spin tune shift could be caused by a small error in the snake current or 
by to some high-order spin perturbation.  As in the case with no Siberian snake in the ring,
the strength of the rf induced resonance determines the tolerable
spin tune shift from the half-integer value (as indicated in Eq.(8)).  
The effect of the
high-order spin perturbation could be reduced by a proper choice of the betatron tunes.
Therefore, an adequate strength of the rf magnet (to achieve $\varepsilon\simeq10^{-3}$) 
and the snake current precision at the level of $10^{-4}$ 
would provide stability of the spin-flipping motion.

In practice, an rf dipole field of 0.1~T could be obtained~\cite{field} in the frequency range
near 20 kHz, which corresponds to a half of the circulation frequency in most high
energy rings.  The spin perturbation strength by a 1-meter-long rf dipole of this type
would be,
\be
\varepsilon =\frac{Ge}{2\pi\, mc^2} \int Bdl \sim 0.01 \ \mbox{(for protons)} \ .
\ee
Such an rf dipole would certainly have adequate strength to control the spin motion.

\section{Spin manipulation of a polarized deuteron beam}

The technique discussed for spin-flipping seems especially attractive for
polarized deuteron beams. The conventional spin manipulation methods become difficult
due to deuteron's small anomalous magnetic moment.  A full Siberian snake 
would only be practical at low energies, where solenoidal magnets could provide 
the required spin rotation (a 200 MeV deuteron beam would require a solenoid with 11~$Tm$ 
field integral).
Similarly, the effect of the rf magnetic field on spin will be rather small,
which limits applicability of the conventional spin-flipping technique. 
In contrast, the method presented here uses the accelerator lattice to flip the spin 
while the rf field keeps the spin motion stable.  This feature becomes an advantage 
for particles with small anomalous magnetic moment in the energy region where 
$G\gamma<10$. In the case of deuteron beam, stable spin-flipping motion could be
organized at the beam energies corresponding to half-integer $G\gamma$ values
\be
T[\mbox{GeV}] = 4.7 + 13.156\cdot n
\ee   
where $n$ is an integer.  A strong rf field could also create stability for the 
non-flipping longitudinal polarization direction when $G\gamma$ is an integer. 
In this case, rf field should be strong enough to dominate over the imperfection 
fields in the accelerator. Note that there are no such
first-order spin perturbations near a half-integer $G\gamma$, and the required rf
field could be much weaker. 

\begin{figure}
\centering\mbox{\epsfysize=3in\epsffile{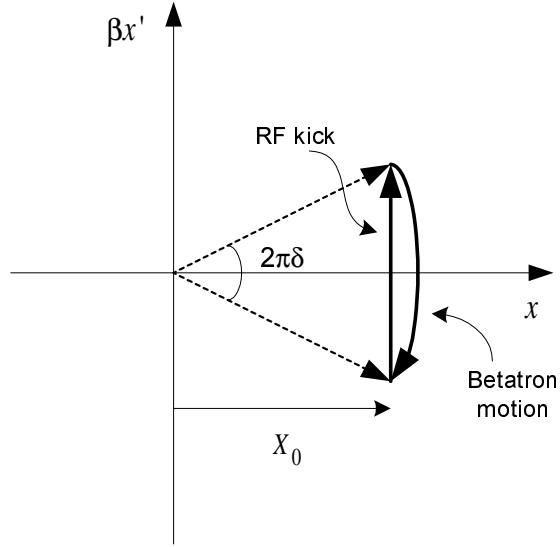}}
\caption{\small Beam oscillations excited by the rf dipole are shown in the phase space 
rotating with the rf frequency. The amplitude of the excited beam oscillations, $X_0$, 
is determined by the rf kick amplitude and the rf frequency separation from the betatron tune 
$\delta=\nu_x-\frac{f_{RF}}{f_{circ}}+k$. This stable mode of the excited beam oscillations is achieved
when rf dipole is turned on adiabatically.}
\end{figure}

The direct effect of the rf magnetic 
field on the deuteron's spin is rather small.  However, an rf-dipole also excite
beam oscillations, and the spin is mainly driven by the quasi-resonant accumulation
of the spin kicks from all quadrupoles of the accelerator lattice.  To estimate the
amplification of the spin perturbation due to the excited beam oscillations, we first
calculate the beam oscillation amplitude using the approach discussed in~\cite{rf-dipole}.  
Considering the beam motion in the phase space coordinates $(x,\, x'\beta_x)$, it is convienient 
to transform to the rf rotating frame where the beam kick by the rf dipole is constant. 
After the rf dipole's kick, the particle's phase space
vector would rotate by an angle,
\be
2\pi\delta=2\pi\left(\nu_x - \frac{f_{RF}}{f_{circ}} + k\right) ,
\ee
where $\nu_x$ is the horizontal betatron frequency and $k$ is an integer. 
When the rf dipole is turned on adiabatically, the excited beam oscillations reach a
stable mode which is shown in Fig.~2. 
The beam oscillation amplitude, $X_0$, for this stable mode
is given by
\be
2\pi\delta\cdot X_0 = \beta_0 x' = \beta_0 \frac{1}{2} \int \frac{B_{RF} d\ell}{B\rho} ,
\ee 
where $\beta_0$ is the value of the horizontal beta-function near the rf dipole. 
The resulting beam oscillations around the ring can be written as
\be
x(s)=X_0\sqrt{\frac{\beta_x(s)}{\beta_0}}\cos\left(\nu_{RF}\theta\right) ,
\label{oscil}
\ee
where $\nu_{RF}=k+ f_{RF}/f_{circ}$ is the rf harmonic closest to the betatron tune 
(i.e. $\nu_{RF}\sim\nu_x$).
The strength of the induced spin perturbation is given by the spin kick accumulated in 
the lattice quadrupoles over one turn around the ring
\be
\varepsilon_b = \frac{G\gamma}{2\pi} \oint g(s) x(s)\, e^{i\nu_s\theta} ds ,
\label{eps_b}
\ee
where $g(s)=\frac{\partial B_y/\partial x}{B\rho}$ is normalized strength of the quadrupoles
around the ring and $\nu_s$ is the spin tune. Eq.(\ref{eps_b}) should be
compared with the strength of the spin perturbation
by the direct effect of the rf magnetic field, which is given by
\be
\varepsilon_0 =\frac{G\gamma}{4\pi} \int \frac{B_{RF} d\ell}{B\rho} .
\ee
In both cases, the effect of transverse magnetic fields is proportional to the anomalous magnetic
moment~\cite{kondrat}.
Substituting Eq.~(\ref{oscil}) into Eq.~(\ref{eps_b}) one would quickly obtain
\be
\varepsilon_b =\varepsilon_0 \frac{\sqrt{\beta_0}}{4\pi\delta} \oint g(s)\sqrt{\beta_x(s)} 
e^{i(\nu_s\pm\nu_{RF})\theta} ds .
\label{e_vs_e}
\ee
When the spin tune is close to the frequency of the induced beam oscillations, $|\nu_s-\nu_{RF}|\ll 1$, 
one can neglect the exponent in the integral.  The remaining integral resembles 
definition of the chromaticity
function $\xi_x = \frac{-1}{4\pi}\oint g(s)\beta_x(s) \simeq -\nu_x$. 
Therefore, the strength of the induced spin perturbation 
can be estimated as  
\be
\varepsilon_b \simeq \varepsilon_0 \frac{\nu_x}{\delta} \frac{\sqrt{\beta_0}}{\langle\sqrt{\beta_x}\rangle} .
\ee
This estimate corresponds to the maximum amplification of the spin perturbation which occurs in the vicinity of the
strong intrinsic depolarizing resonances $\nu_s\simeq\nu_x$, where the spin kicks from all lattice 
quadrupoles are synchronized with the spin precession. Farther away from these regions, the effect of the
induced beam oscillations on spin is reduced by the exponent in the Eq.~(\ref{e_vs_e}); in that
case, smaller $\delta$ would enhance the excited beam oscillation amplitude as well as the induced
spin perturbation. 

Finally, we would like to comment that acceleration of polarized deuterons 
seems possible in modern high energy rings~\cite{courantD}.
While full Siberian snakes do not seem practical for high energy deuterons, their 
depolarizing resonances are 25 times weaker and 25 times farther apart than for protons. 
Therefore, one could use individual resonance correction
techniques developed for proton beam at the AGS ring (Brookhaven Natl. Lab.).  
A partial Siberian snake could overcome all imperfection depolarizing resonances~\cite{partial}.
Such a partial snake could be realized either using a solenoid magnet or a set
of correction dipoles distributed around the ring to create a controlled closed 
orbit perturbation.  Note that, in an ideal case, reversing the axis of the partial
snake could flip the longitudinal polarization of the beam stored near an integer $G\gamma$.
However, without additional correction of the natural imperfection resonance 
this spin-flipping method remains impractical due to inevitable polarization losses.
The intrinsic depolarizing resonances could be handled by
an rf dipole which induces coherent beam oscillations and
makes intrinsic resonances strong enough to spin-flip~\cite{dipole-ags}.   
An rf spin perturbation could also be used as a spin rotator for polarized deuterons. 
For example, to bring initial vertical beam polarization to the longitudinal direction,
a horizontal rf dipole field could be applied for a time period which corresponds to 
the required $\pi/2$ spin rotation.

\section{Summary}

In summary, we found that an external rf magnetic field can be used to
create a stable mode of the spin motion, where the polarization direction
flips after every turn around the ring.  Such stable spin-flipping spin
motion can be realized whenever the spin tune is equal to a half integer
value; this is always true in the rings equipped with Siberian snakes, while 
rings without Siberian snakes also reach half-integer spin tune at certain energies.    
The applied rf magnetic field should be orthogonal to the 
unperturbed stable spin direction, and should operate at exactly one half
of the circulation frequency. Provided that the rf field is synchronized with
the circulating beam and is strong enough to dominate over possible spin tune
spread, this spin-flipping motion is stable. 
It was also noted earlier~\cite{rf-derb}, that 
the rf stabilization of the spin motion against the spin tune spread
is an interesting possibility for accelerators with Siberian snakes.



\begin{thebibliography}{99}
\bibitem{iucf-flip}
D.D. Caussyn {\it et al.}, Phys. Rev. Lett. {\bf 73}, 2857 (1994);\\
B.B. Blinov {\it et al.}, Phys. Rev. Lett. {\bf 81}, 2906 (1998).
\bibitem{spinor}
B.W. Montague, Part. Accel. {\bf 11}(4), 219 (1981).
\bibitem{snk-idea} 
Ya.S. Derbenev and A.M. Kondratenko, Sov. Phys. Dokl. {\bf 20}, 562 (1978).
\bibitem{courant-lee}
S.Y. Lee and E.D. Courant, Phys. Rev. {\bf D 41}, 292 (1990). 
\bibitem{snk-res}
S.Y. Lee and S. Tepikian, Phys. Rev. Lett. {\bf 56}, 1635 (1986);\\
R.A. Phelps {\it et al.}, Phys. Rev. Lett. {\bf 78}, 2772 (1997).
\bibitem{field}
B. Parker {\it et al.}, in Proc. of 1999 Part. Accel. Conf. (PAC-99, New York), 3336 (1999);
P. Schwandt, private communications.
\bibitem{rf-dipole}
M. Bai {\it et al.}, Phys. Rev. {\bf E 56}, 6002 (1997). 
\bibitem{kondrat}
A.M. Kondratenko, in Proc. of 9$^{th}$ Intl. Symposium on High Energy Spin Physics,
Bonn 1990, 140 (1991).
\bibitem{courantD} 
E.D. Courant, in Proc. of Workshop on RHIC Spin Physics, BNL April 1998,
Report BNL-65615, 275 (1998); Spin Note AGS/RHIC/SN 066 (1997).
\bibitem{partial}
V.A. Anferov {\it et al.}, Phys.\ Rev. {\bf A46}, R7383 (1992);\\
H. Huang {\it et al.}, Phys.\ Rev.\ Lett.\ {\bf 73}, 2982 (1994).
\bibitem{dipole-ags}
M. Bai {\it et al.}, Phys. Rev. Lett. {\bf 80}, 4673 (1998). 
\bibitem{rf-derb}
Ya.S. Derbenev, in Proc. of DESY Workshop on Polarized Protons at High Energies,
DESY-PROC-1999-03, 225 (1999).
\end{thebibliography}
\end{document}